\documentclass[fleqn,usenatbib]{mnras}

\usepackage{newtxtext,newtxmath}
\usepackage[T1]{fontenc}
\usepackage{ae,aecompl}
\usepackage[table]{}

\usepackage{graphicx}	% Including figure files
\usepackage{amsmath}	% Advanced maths commands
\usepackage{amssymb}	% Extra maths symbols
\usepackage{hyperref}
\usepackage{tablefootnote}
\usepackage{wasysym}
\usepackage{float}
\usepackage{tabularx}
\usepackage{comment}
\title[Discovery of GPX-TF16E-48]{Discovery of a pre-cataclysmic binary with unusual chromaticity of the eclipsed white dwarf by the GPX Survey}

\author[Krushinsky et al.]{
Vadim Krushinsky$^{1}$\thanks{E-mail: krussh@gmail.com},
Paul Benni$^{2}$,
Artem Burdanov$^{3,4}$,
Igor Antokhin$^{5}$,
\newauthor
Eleonora Antokhina$^{5}$,
Emmanu{\"e}l Jehin$^{6}$,
Khalid Barkaoui$^{7,8}$,
Alan Fitzsimmons$^{9}$,
\newauthor
Christopher Gibson$^{9}$,
Micha{\"e}l Gillon$^{7}$,
Alexander Popov$^{1}$,
\"Ozg\"ur Ba\c{s}t\"urk$^{10}$,
\newauthor
Zouhair Benkhaldoun$^{8}$,
Alessandro Marchini$^{11}$,
Riccardo Papini$^{12}$,
Fabio Salvaggio$^{12}$,
\newauthor
Varvara Brazhko$^{1}$
\newauthor
\\
$^{1}$Ural Federal University, 620002, Mira Street, 19, Yekaterinburg, Russian Federation\\
$^{2}$Acton Sky Portal (Private Observatory), Acton, MA, USA\\
$^{3}$Department of Earth, Atmospheric and Planetary Sciences, Massachusetts Institute of Technology, 77 Massachusetts Avenue, Cambridge, MA 02139, USA\\
$^{4}$Instituto de Astrof\'isica de Canarias, V\'ia L\'actea s/n, 38205 La Laguna, Tenerife, Spain\\
$^{5}$Moscow Lomonosov State University, Sternberg State Astronomical Institute, 119992 Universitetsky prospect, 13, Moscow, Russian Federation\\
$^{6}$Space Sciences, Technologies and Astrophysics Research (STAR) Institute, Universit\'e de Li\`ege, All\'ee du 6 Ao\^ut 19C, 4000 Li\`ege, Belgium\\
$^{7}$Astrobiology Research Unit,  Universit\'e de Li\`ege, All\'ee du 6 Ao\^ut 19C, 4000 Li\`ege, Belgium\\
$^{8}$Oukaimeden Observatory, High Energy Physics and Astrophysics Laboratory, Cadi Ayyad University, Marrakech, Morocco\\
$^{9}$Astrophysics Research Centre, Queen's University Belfast, Belfast BT7 1NN, Northern Ireland\\
$^{10}$Ankara University, Faculty of Science, Department of Astronomy and Space Science, TR-06100 Tandogan, Ankara, Turkey\\
$^{11}$Astronomical Observatory - DSFTA, University of Siena, Via Roma 56, 53100 Siena, Italy\\
$^{12}$Wild Boar Remote Observatory, San Casciano in Val di Pesa (FI), Italy\\
}

\date{Accepted XXX. Received YYY; in original form ZZZ}

\pubyear{2019}

\begin{document}
\label{firstpage}
\pagerange{\pageref{firstpage}--\pageref{lastpage}}
\maketitle

\begin{abstract}
We report the discovery of a relatively bright eclipsing binary system, which consists of a white dwarf and a main sequence K7 star with clear signs of chromospheric and spot activity. The light curve of this system shows $\sim0.2$\,mag ellipsoidal variability with a period of 0.297549\,d and a short total eclipse of the white dwarf. Based on our analysis of the spectral and photometric data, we estimated the parameters of the system. The K7V star is tidally deformed but does not fill its Roche lobe (the filling factor is about 0.86). The orbital inclination is $i=73^\circ.1\pm 0^\circ.2$, the mass ratio is $q=M_2/M_1\approx 0.88$. The parameters of the K7V star are $M_2\approx 0.64$\,M$_{\odot}$, $R_2=0.645\pm 0.012$\,R$_{\odot}$,  $T_2\approx 4070$\,K. The parameters of the white dwarf are $M_1\approx 0.72$\,M$_{\odot}$, $R_1=0.013\pm 0.003$\,R$_{\odot}$,  $T_1=8700\pm 1100$\,K. Photometric observations in different bands revealed that the maximum depth of the eclipse is in the \textit{SDSS r} filter, which is unusual for a system of a white dwarf and a late main sequence star. We suspect that this system is a product of the evolution of a common envelope binary star, and that the white dwarf accretes the stellar wind from the secondary star (the so-called low-accretion rate polar, LARP).
\end{abstract}


%%%%%%%%%%%%%%%%%%%%%%%%%%%%%%%%%%%%%%%%%%%%%%%%%%

%%%%%%%%%%%%%%%%% BODY OF PAPER %%%%%%%%%%%%%%%%%%

\section{Introduction}\label{Introduction}

A stellar binary, which consists of a white dwarf (WD) primary and a main sequence (MS) secondary star is referred to as WDMS and it is a result of the evolution of the system of two MS stars, where one of the companions turns into the WD. A WDMS system evolves from an initially wider binary system, where the primary companion evolves into a red giant forming a common envelope containing the less massive and, therefore, less evolved secondary companion \citep{2004A&A...419.1057W}. During the common envelope phase, stars transfer orbital angular momentum to the surrounding matter, what causes decrease of the orbital separation between them. Later, the common envelope is ejected from the system and a WDMS system is formed \citep{1976IAUS...73...75P}. Such systems are denoted as post common envelope
binaries (PCEB). If components of the WDMS have small orbital separation, gravity of the WD distorts the secondary companion. As a result, one can observe an ellipsoidal modulation of the light curve due to the  change of the observed area of the secondary. Other periodical modulations occur because of the temperature gradient of the secondary due to gravitational darkening, absorption and reflection of radiation emitted by the WD. Further angular momentum loss occurs due to various mechanisms (magnetic braking, gravitational wave radiation, etc.), and separation decreases until the secondary overflows Roche lobe and the WDMS turns into a cataclysmic variable (e.g., \citealt{2011ApJS..194...28K}). When the projected distance between the centres of the companions (impact parameter) is favourable, then eclipses might be observed. In this case, a depth of the eclipse depends on a size of the WD and its temperature relative to the secondary companion. In a simple black-body model, if the WD is hotter than the MS, the eclipse depth monotonically increases from IR to UV (e.g., see SDSS J030308.35+005444.1 light curve from \citealt{2013MNRAS.436..241P}).

WDMS systems are favourable objects for observations and studies, as possible progenitors of cataclysmic variables and as benchmarks for testing stellar evolutionary models. Eclipsing WDMS systems make it possible to infer the absolute sizes and masses of companions and also make it possible to test the WD mass-radius relationship \citep{2017MNRAS.470.4473P}. Thanks to the extensive modern surveys like SDSS \citep{2000AJ....120.1579Y} and CRTS \citep{2009ApJ...696..870D}, numerous eclipsing WDMS systems have been discovered (e.g., \citealt{2013MNRAS.429..256P,2015MNRAS.449.2194P}), but eclipsing systems still represent a minor part of the known population of WDMS systems \citep{2018MNRAS.477.4641R}. Exoplanet transit surveys, in turn, also expand the current population of eclipsing WDMSs. For example, \cite{2014MNRAS.437.1681M} reports on the discovery of 17 new eclipsing WDMS systems from the ground-based WASP survey \citep{Pollacco2006} and \cite{2015ApJ...815...26F} describes discoveries of new short-period eclipsing systems with low-mass WDs from \textit{Kepler} light curves \citep{2010Sci...327..977B}.

Based on the prototype KPS survey \citep{2016MNRAS.461.3854B}, the Galactic Plane eXoplanet survey (GPX, \citealt{2017JAVSO..45..127B,2018PASP..130g4401B}) performs photometric observations of the Galactic plane to search for new transiting exoplanets. However, all the stars are routinely scanned for any signs of variability. In this paper, we present the discovery of GPX-TF16E-48 variability in the GPX wide-field data and attempt to explain the peculiar photometric behaviour of this eclipsing WDMS system using additional photometric and spectroscopic observations. 

The rest of the paper is structured as follows: Section~\ref{Observations} describes GPX discovery wide-field photometry, subsequent follow-up observations with different instruments and data reduction. Section~\ref{ANALYSIS} is devoted to the light curve and spectral analysis. In Section~\ref{sec_disc} we discuss the results obtained and make assumptions about the nature of this eclipsing binary. In the Section~\ref{CONCL} we outline our findings.

\section{DISCOVERY AND FOLLOW-UP OBSERVATIONS}\label{Observations}

\subsection{GPX detection photometry and open source data}
\begin{figure*}
  \includegraphics[width=\textwidth]{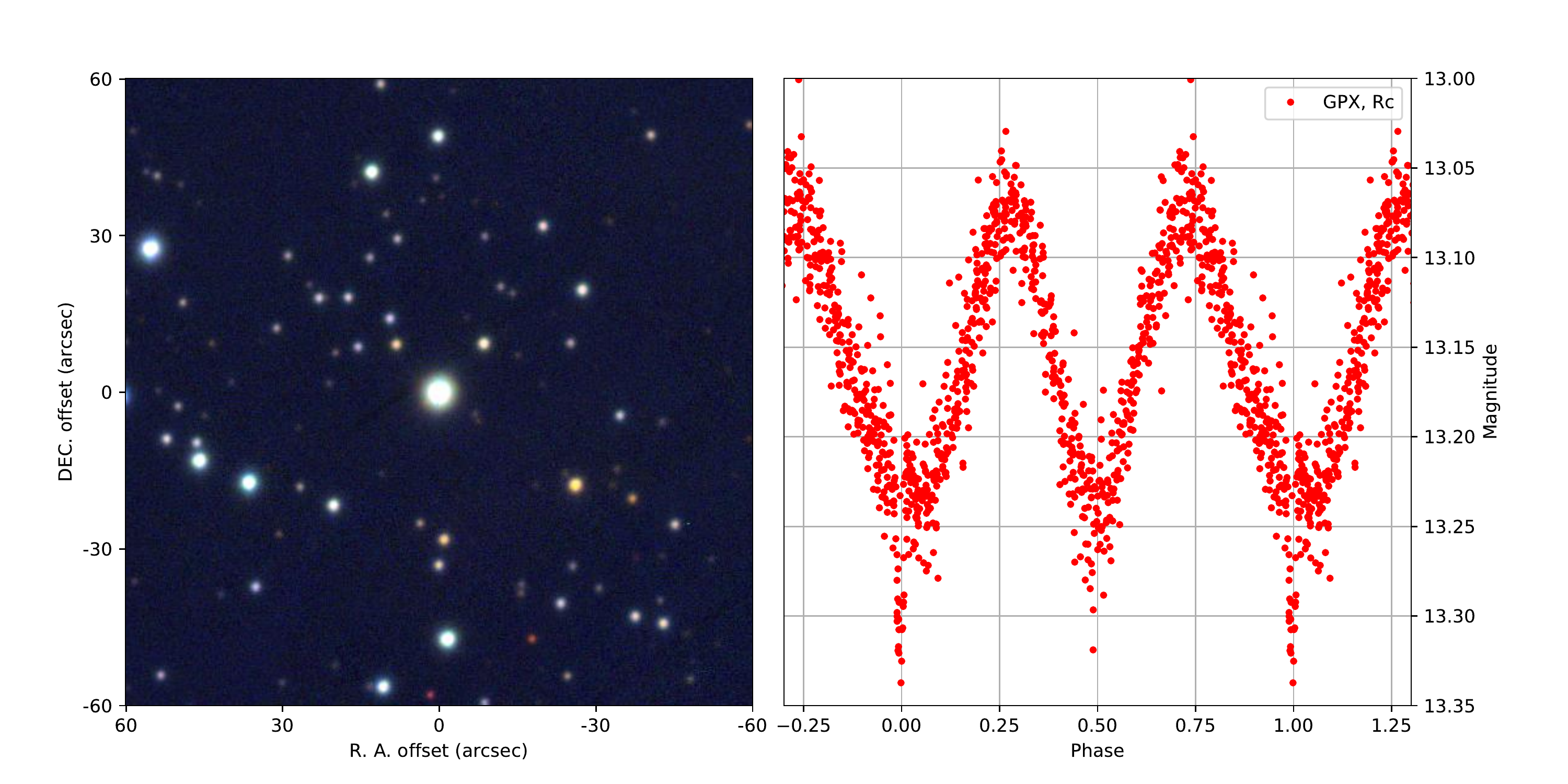}
  \caption{Left: PanStarrs-1 colour (i/r/g) image of the 2${^\prime}$x2${^\prime}$ area around GPX-TF16E-48. Right: GPX discovery light curve folded with the 7.14\,h period.}
    \label{fig_Map_and_DLC}
\end{figure*}

The TF16E field of the GPX survey was observed with the RASA telescope from 2018 August 5 to 2018 October 22 for 27 nights in the $R_c$ filter. The RASA telescope is a Rowe-Ackermann Schmidt Astrograph (0.28-m, f/2.2) wide-field telescope and it is the main instrument of the GPX wide-field exoplanet search. It is based at the Acton Sky Portal private observatory in Acton (MA, USA) and it was built from readily available off-the-shelf equipment. GPX-TF16E-48's photometric light curve was automatically selected as a possible variable star using our automatic reduction pipeline. The pipeline includes standard photometric and astrometric \citep{Lang_2010} reductions, flux extraction using {\sc IRAF} \citep{1993ASPC...52..173T}, differential photometry using close ensemble of comparison stars and search for objects with significant variability. Identification of variable stars is based on the RoMS (Robust Median Statistics) criterion, described in \cite{Rose_2007}. A detailed description of the pipeline is provided in \citealt{2014AstBu..69..368B, 2016MNRAS.461.3854B}. Indeed, the target's light curve shows variability, typical to W~Ursae Majoris-type eclipsing variables (EW) with an amplitude of 0.175\,mag and a period of 7.14\,h ($\sim 0.3$\,d). However, the light curve had a few surprises: a box-shaped eclipse with an amplitude of 0.06\,mag is seen and the positions of the maxima of the main 0.175\,mag variability are slightly shifted from $-0.25$ and $+0.25$ phases. The discovery light curve folded with the 7.14\,h period is shown in Fig.~\ref{fig_Map_and_DLC}, where the occurring box-shaped eclipse has 0.00 phase.

The Gaia DR2 parallax of the object is 5.137$\pm0.0194$\,mas, which corresponds to a distance of 194.7$\pm0.7$\,pc  \citep{2018A&A...616A...1G}. The colour indices and apparent magnitudes from different catalogues match with a K7-M1 dwarf located at 100-200\,pc. There are several magnitude measurements from Pan-STARRS PS2 \citep{2016arXiv161205560C}, which are in agreement with the observed variability from the GPX survey. GPX-TF16E-48 is a faint UV source in the NASA/IPAC Extragalactic Database (NED), and it is not presented in any X-ray database. The main variability of the object was independently detected by Vitali Nevski using ASAS-SN database and added to the VSX database on 2018 December 14 as an EW variable star\footnote{\url{https://www.aavso.org/vsx/index.php?view=detail.top\&oid=629066}}. 

We summarize all the information about GPX-TF16E-48 in Table~\ref{Tab1_Info} and provide a finding chart in Fig.\ref{fig_Map_and_DLC}

\begin{table}
\centering 
\caption{General information about GPX-TF16E-48. B and V magnitudes are from the APASS catalog (\citealt{2015AAS...22533616H}), J magnitude -- from the 2MASS catalog (\citealt{2006AJ....131.1163S}).}
\begin{tabular}{ll}
\hline
Identifiers & GPX-TF16E-48\\
& Gaia~DR2~531486745299867904\\ 
& 2MASS~01294349+6715300\\
& VSX NEV239\\
RA (J2000) & 01h 29m 43.488s\\ 
DEC (J2000) & +67$^{\circ}$ 15${^\prime}$ 30.02${^{\prime\prime}}$\\
Gaia parallax & 5.137$\pm0.0194$~mas\\
Bmag & 15.936 \\
Vmag & 14.713\\
Jmag & 11.953\\
Period & $0.297549\pm0.000001$\,d\\
Epoch(HJD) & $2458762.2285\pm0.0005$\\
\hline
\label{Tab1_Info}
\end{tabular}

\label{tab:general_info}
\end{table}

\subsection{Follow-up observations}\label{follow-up_obs}
\begin{figure*}
  \includegraphics[width=\textwidth]{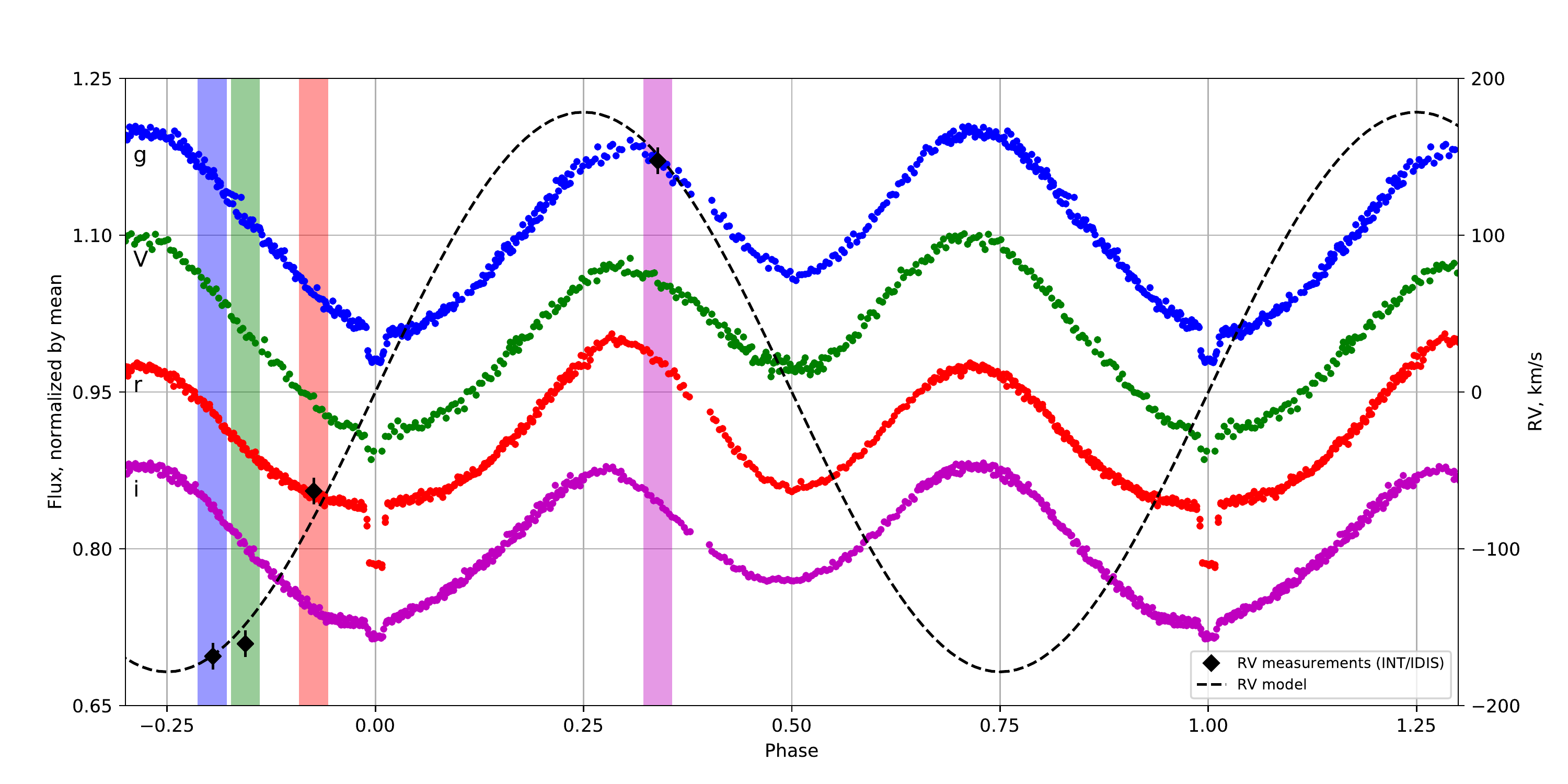}
  \caption{Period-folded light curves of GPX-TF16E-48 (left axis) and radial velocity measurements (right axis). Photometric data obtained on 2019 April with the Kourovka 1.2-m telescope (see Subsection~\ref{follow-up_obs} for details). All light curves were normalised and shifted vertically by 0.1 for visual clarity. Colours of vertical stripes corresponded to colours of spectra on Fig.~\ref{fig_Spec}, and their widths are equal to exposure time.}
  \label{fig_RV_LC} 
\end{figure*}

Photometric follow-up observations of GPX-TF16E-48 were carried out with seven telescopes at five observatories (see Table~\ref{TblObs} for a summary). Short time-series before, during and after the eclipse, and during the opposite phase were obtained with the TRAPPIST-North telescope \citep{2017JPhCS.869a2073B,2011Msngr.145....2J,2011EPJWC..1106002G} in November and December 2018 in the B, V, $R_C$, narrow-band H$\alpha$ and I+z bands. High precision observations of the eclipse in the B filter were obtained on 2018 November 30 with the SAI MSU 2.5-m telescope \citep{2017ARep...61..715P}. Simultaneous observations covering the full period were obtained using the Kourovka 1.2-m telescope in the V band on 2019 April 4 and in the \textit{SDSS g, r} and \textit{i} bands on 2019 April 9-10 (Fig.~\ref{fig_RV_LC}) and October 7 using a three-channel CCD photometer.

The T\"{U}B\.{I}TAK National Observatory T100 telescope, MASTER-Ural~\citep{2010AdAst2010E..30L} and Acton Sky Portal 14${^{\prime\prime}}$SCT telescopes observed GPX-TF16E-48 for several nights in 2018 and 2019 in different filters (see Table~\ref{TblObs}).

Photometric reductions were performed in a standard way using dark frames and twilight sky flat-field images. Fluxes were extracted with the aperture photometry technique using fixed aperture radius $\sim$1.5\,FWHM, where FWHM is the mean full width at half maximum of the stellar point spread function (PSF) during the observing run. We used an ensemble of nearby non-variable stars with similar to the GPX-TF16E-48 colour indices to derive the target's differential light curve and we checked all reference stars light curves for possible correlations with airmass, pixel position, FWHM and sky background.

\begin{table*}
\begin{center}
\caption[]{Follow-up observations log}\label{TblObs}
\begin{tabularx}{1.0\textwidth}{lllll}
%\begin{tabular}{lllll}
\hline\noalign{\smallskip}
Observatory &Telescope &Date &Filters/Dispersion &Comments\\
\hline\noalign{\smallskip}
Acton Sky Portal &RASA, 0.28-m, f/2.2 &Aug.-Oct. 2018 &$R_c$ &22 nights, discovery light curve\\
Acton Sky Portal &Celestron, 0.35-m, f/11 &Nov.-Dec. 2018 &\textit{g, r, i} &6 nights, follow-up observations\\
Ouka\"imeden & TRAPPIST-North, 0.6-m, f/8 &Nov.-Dec. 2018 &B, V, $R_c$, I+z, H$\alpha$ &5 observations of the eclipse\\
& & & & in different bands\\
Kourovka &1.2-m, f/10 &Apr., Oct. 2019 &V, \textit{g, r, i} & 3+1 nights of high S/N observations;\\ 
& & & & simultaneous observations in \textit{g, r} and \textit{i}\\
Kourovka &MASTER, 0.4-m, f/2.5 &Jan.-Apr. 2019 &V, R &18 nights; simultaneous observations\\ 
& & & & in V and R\\
T\"{U}B\.{I}TAK &T100, 1.0-m, f/10 &18 Feb. 2019 &I &Observations of one period\\
CMO SAI MSU &2.5-m, f/10 &30 Nov. 2018 &B &High S/N observation of the eclipse\\
Roque de los Muchachos &INT/IDS, 2.5-m &14 and 17 Dec. 2018 &3800--6700\,\r{A}, 1.5\,\r{A}/pix. &4 low resolution spectra\\
\noalign{\smallskip}\hline
%\end{tabular}
\end{tabularx}
\end{center}
\end{table*}

\begin{figure*}
  \includegraphics[width=\textwidth]{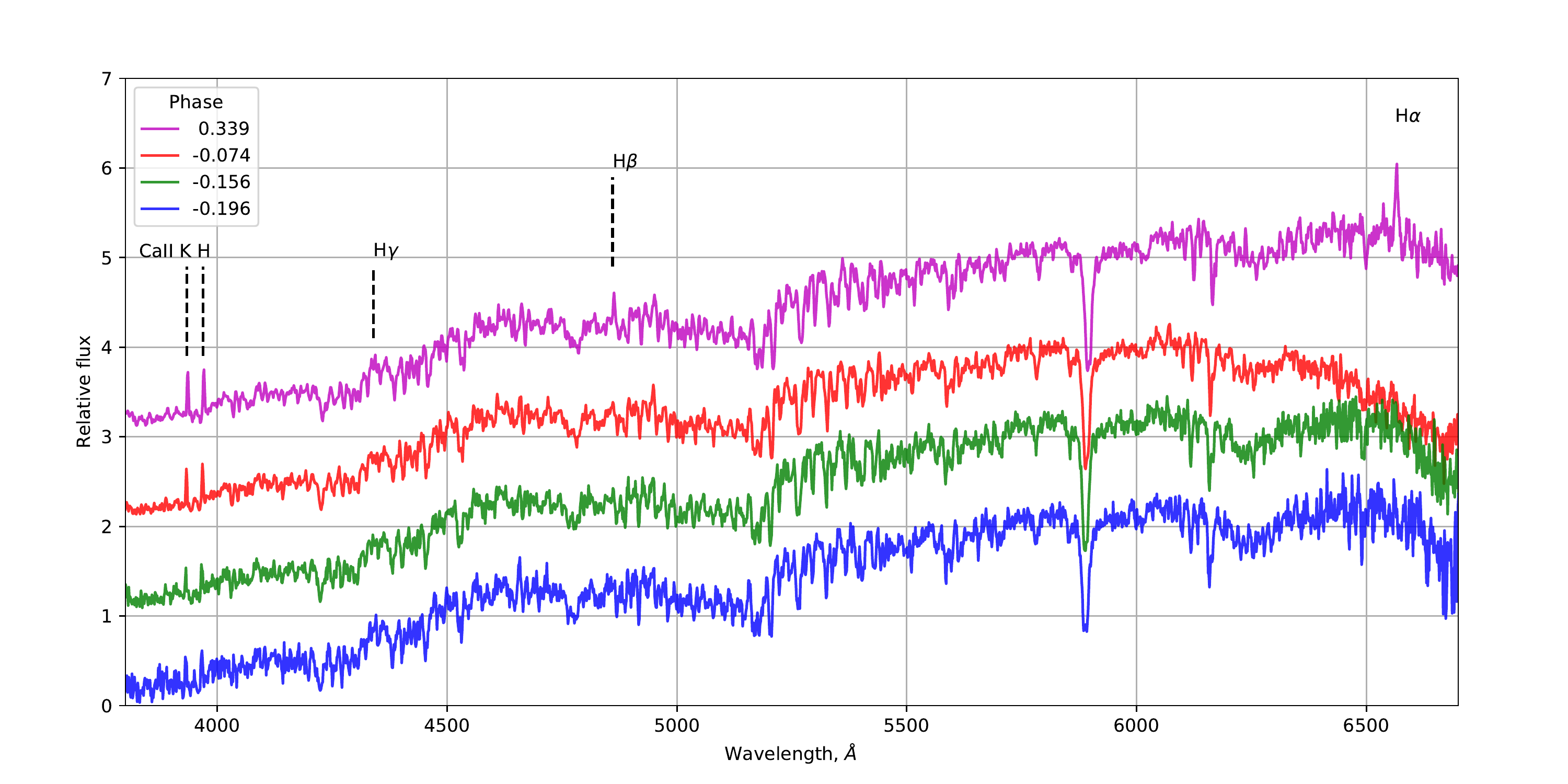}
  \caption{Four spectra were obtained with the Isaac Newton Telescope/Intermediate Dispersion Spectrograph on 2018 December 14 and 17. Strong Ca\,\textsc{II} H-K emissions are present in all spectra. The Balmer lines are clearly visible on phase 0.339. All spectra were obtained at same zenith distance (43--41$^{\circ}$) but in different weather and seeing condition (see Subsection~\ref{follow-up_obs} for details). Spectra were normalized to the mean flux in 4000--5000\,\r{A} region and shifted vertically for visual clarity. The bend of continuum near 6500\,\r{A} on phase $-0.074$ is due to data reduction issues.}
  \label{fig_Spec}
\end{figure*}

%\subsection{Spectroscopic follow-up observations}
Four spectra of the GPX-TF16E-48 star were obtained with the Isaac Newton Telescope/Intermediate Dispersion Spectrograph (INT/IDS) on 2018 December 14 and 17. These observations correspond to phases ${-0.196}$, ${-0.156}$, ${-0.074}$ and ${0.339}$ (Fig.~\ref{fig_Spec}). All exposures were 900 seconds long and a slit width of 1.0${^{\prime\prime}}$ was used. The seeing varied for different exposures and on 2018 December 17 there was a considerable amount of dust in the atmosphere. The spectra were reduced using the standard INT/IDS pipeline. The wavelength calibration is accurate to 0.1\r{A} (rms) or better for all spectra. The pixel scale in the data is $\sim$1.5\r{A} per pixel, so wavelength calibration is better than 0.1 pixel. Radial velocities measurements are provided in Table~\ref{TblRV}.

\begin{table}
\begin{center}
\caption[]{Measured radial velocities of K7V companion}\label{TblRV}
\begin{tabular}{lrr} %{1.0\columnwidth}
\hline\noalign{\smallskip}
Obs. midpoint &Phase &Barycentric RV\\
(BJD) & &(km\,s$^{-1}$)\\
\hline\noalign{\smallskip}
2458467.29929651 &-0.196 &$-205\pm9$\\
2458467.31110172 &-0.156 &$-197\pm9$\\
2458470.31100902 &-0.074 &$-99\pm9$\\
2458470.43392171 & 0.339 &$ 111\pm9$\\ 
\noalign{\smallskip}\hline
\end{tabular}
\end{center}
\end{table}

Photometric and spectroscopic data are available for download from the Kourovka Observatory file sharing server.\footnote{\url{https://optlab.kourovka.ru/GPX_TF16_48/}}

\section{DATA ANALYSIS}\label{ANALYSIS}

Ephemeris of GPX-TF16E-48 were determined using a period search service \textsc{Webefk}\footnote{\url{http://vast.sai.msu.ru/lk/}} based on the Lafler-Kinman method~\citep{1965ApJS...11..216L} and are given in Table~\ref{Tab1_Info}.

%Spectra
Spectral classification of the red dwarf component of GPX-TF16E-48 was done by comparing the spectra with the spectral library \citep{Valdes}. We determined the spectral class as K7V, which is in agreement with the Gaia parallax. Strong Ca\,\textsc{II} H-K emissions are present in all spectra, indicating significant chromospheric activity of the MS companion \citep{1968ApJ...153..221W}. Balmer emission lines are clearly visible on the phase 0.339. Velocities of the emission lines correspond to the red dwarf companion. We did not detect any obvious spectral signs of the WD in the spectra. A library spectrum of HD237903 was used as a template for cross-correlation measurements of the radial velocities. The semi-amplitude of radial velocity is $K_2 = 178\pm10$\,km\,s$^{-1}$ and it was obtained under the assumption of a circular orbit.

\begin{figure*}
  \includegraphics[width=\textwidth]{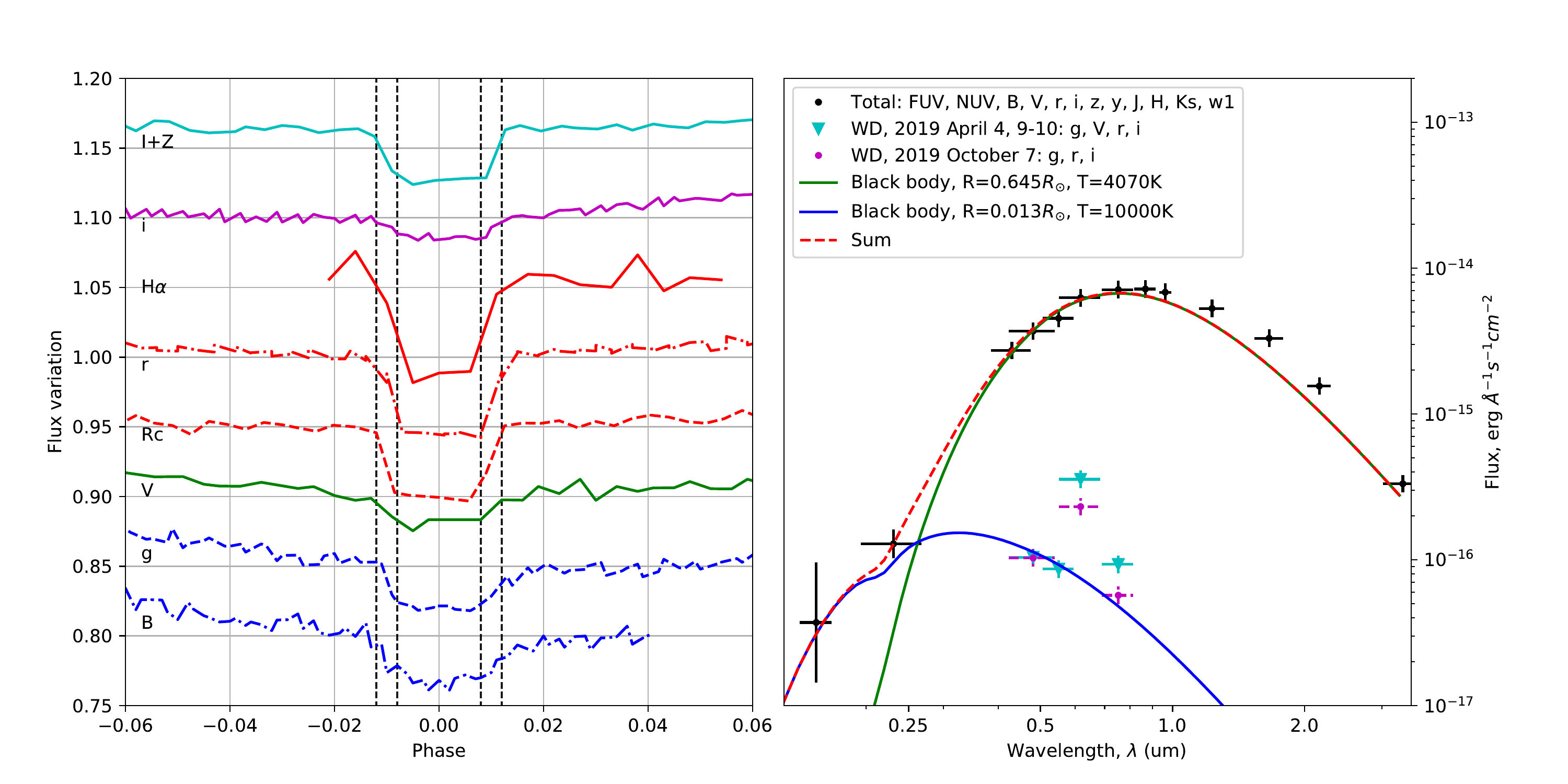}
  \caption{The WD eclipse in various bands shown on the left panel. From TRAPPIST-North: I+Z on 2018 December 5; H$\alpha$ on 2018 December 12; $R_C$ on 2018 December 10. From SAI MSU 2.5-m telescope: B on 2018 November 30. From Kourovka 1.2-m telescope: V on 2019 April 4; \textit{SDSS g, r, i} on 2019 April 9 and 10. The ingress and egress of the eclipse is $\sim$100s long, full duration of the eclipse is $\sim$600s. 
  The {\em GALEX} FUV and NUV, APASS B and V, PS2 \textit{g, r, i, z} and \textit{y}, 2MASS J, K and Ks, {\em WISE} w1 spectral points are shown on the right panel. Also shown are the fluxes of the eclipsed body, calculated from the depth of the eclipse for two observation sets in April and October 2019. 
  The green and blue curves are black body models for the WD and MS components at 200\,pc and \textit{$A_{V}=0.31$}.}
    \label{fig_eclipse}
\end{figure*}

%eclipse geometry and eclipse photometry \comment{Kepler}
The ingress and egress of the total eclipse are ${\sim}$\,100\,s long and the full duration of the eclipse is  only 600\,s (Fig.~\ref{fig_eclipse}). Based on the shape of the light curve and total eclipse, we assumed that the system consists of a MS star with a large Roche lobe filling factor and a WD. The maximum temperature of the WD should not exceed $10\,000$\,K from the low flux in the {\em GALEX} FUV and NUV and the depths of the eclipse in different bands (Fig.~\ref{fig_eclipse}, right panel).

\subsection{Light curves modelling}\label{sec_modelling}

\begin{figure*}
  \includegraphics[width=\textwidth]{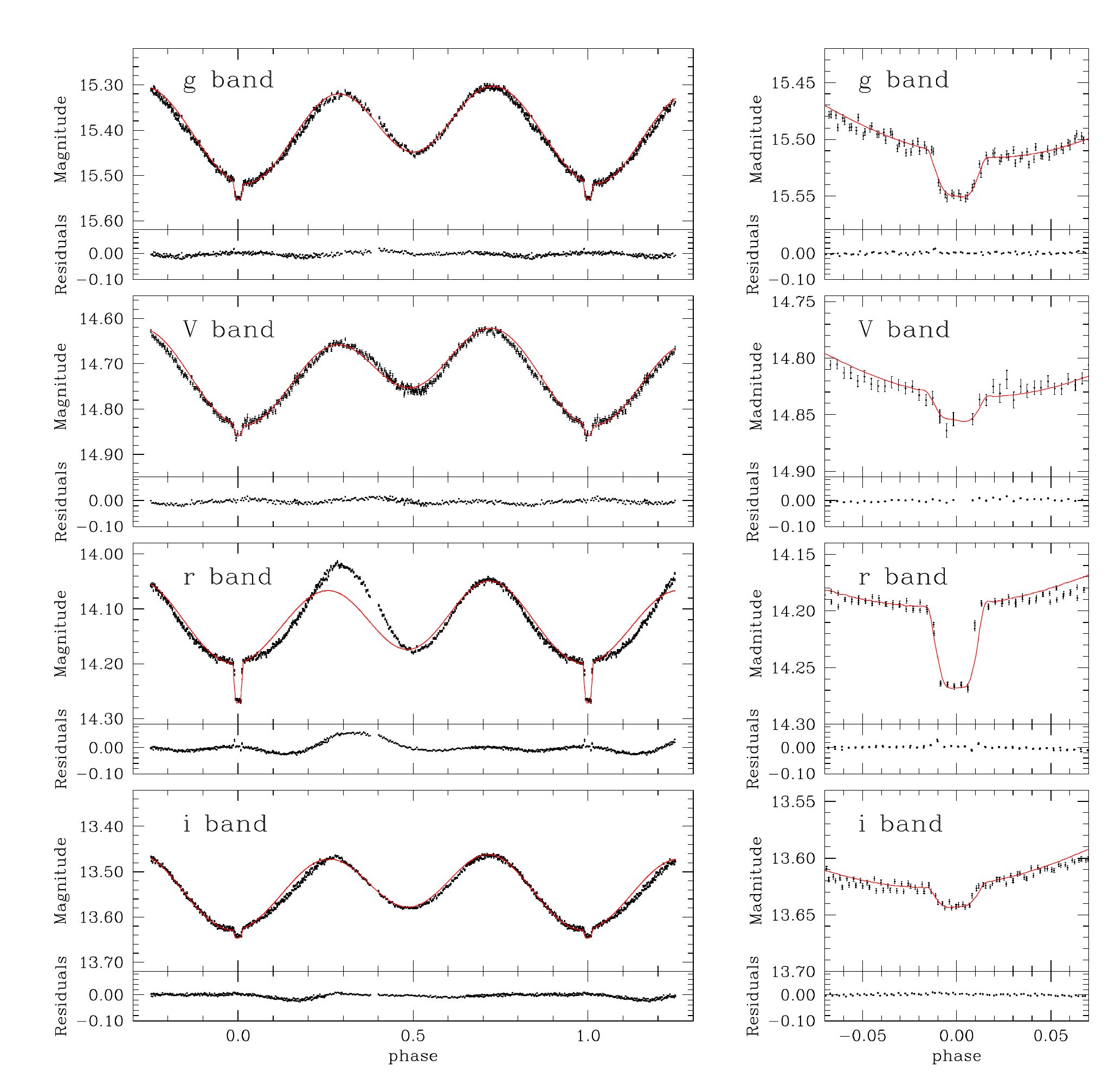}
  \caption{The light curves in the four analysed bands are shown in the left panel. The eclipse part of the light curves are in the right panel. The photometric parameters of the model solution are presented in Tables~\ref{tab_modelpars},\,\ref{tab_abspars}. The observed light curve in the \textit{r} band has abnormal shape. The model light curve in this band is computed with geometric parameters set to those found in the \textit{g} band. $T_1$ was obtained by fitting the depth of the total eclipse. For details, see the text.}
    \label{fig_lcfit}
\end{figure*}

The light curves of GPX-TF-16E-48 clearly show significant ellipsoidal variability and a short total eclipse of the compact object with rather small amplitude. Similar light curves have been observed in RR Caeli \citep{1998AJ....116..908B} and SDSS 0303+0054 \citep{2009MNRAS.394..978P}, which were classified as post-common envelope binaries consisting of a WD and a MS star. The ellipsoidal variability of RR~Caeli and SDSS 0303+0054 is notably smaller than that of \mbox{GPX-TF-16E-48} ($\sim0.2$\,mag), which clearly indicates that in the latter case, the MS star is strongly deformed and (nearly) fills its critical Roche lobe. To estimate the stellar and orbital parameters of the system, we performed an analysis of four light curves, fully covering the orbital period, and observed with the 1.2-m Kourovka telescope: V band (2019 April 4) and \textit{SDSS g}, \textit{r}, and \textit{i} bands (2019 April 9). The amplitudes of the total eclipse of the WD are equal to $\sim0.04$\,(\textit{g}), $\sim0.03$\,(V),  $\sim0.07$\,(\textit{r}), $\sim0.03$\,mag (\textit{i}) (see Fig.~\ref{fig_lcfit}, right panels). The light curves are notably asymmetric, the fluxes of the two maxima are different, the ingress and egress parts of the primary ellipsoidal minimum are also asymmetric. Also, if only ellipsoidal variability was present, its depth around the orbital phase 0.0 would be smaller than that at the phase 0.5. These features indicate the presence of spot(s) on the surface of the MS star \citep{2007A&A...474..205T, 2016MNRAS.458.2793P, 2019AJ....157....3L}. 

\begin{table*}
\centering
 \caption{Photometric solutions.}
 \label{tab_modelpars}
 \begin{tabular}{lccccl}
 \hline
\noalign{\smallskip}
 Parameter & \textit{g} &  V & \textit{r}$^{\mathrm{a}}$ & \textit{i} & Parameter \\
           &   &    &   &   & status \\
 
 \noalign{\smallskip}
 \hline 
 \noalign{\medskip}

 Mass ratio $q=M_2/M_1$ &  $0.885$ &  $0.885$ & $0.885$ & $0.885$ & adopted \\
 Inclination $i$ ($^{\circ}$)  &  $73.10 \pm 0.20$  & $73.15 \pm 0.20$ & $73.10$ & $73.00 \pm 0.20$  & adjusted \\ 
 Roche lobe filling factor $\mu_1$ & $0.017 \pm 0.002$ & $0.016 \pm 0.002$ & $0.017$ & $0.018 \pm 0.002$  & adjusted \\
 Roche lobe filling factor $\mu_2$ & $0.860 \pm 0.010$ & $0.865 \pm 0.011$ & $0.860$ & $0.862 \pm 0.010$  & adjusted \\
 Surface potential $\Omega_1$   & $161$ & $171$ & $161$ & $153$   & computed \\
 Surface potential $\Omega_2$   & $3.991$ &  $3.973$ & $3.991$ & $3.984$  &  computed \\
 Temperature of primary $T_1$\,(K)   & $8600 \pm 1000$ & $8800 \pm 1000$  & $22\,000 \pm 1300$ & $14\,000 \pm 1000$  & adjusted \\ 
 Temperature of secondary $T_2$\,(K) & $4070$ & $4070$ & $4070$ & $4070$  & adopted \\
 $L_1/(L_1+L_2)^{\mathrm{b}}$ & $0.029$ & $0.016$ & $0.056$ & $0.012$  & computed \\
 $L_2/(L_1+L_2)^{\mathrm{b}}$ & $0.971$ & $0.984$ & $0.944$ & $0.988$  & computed \\
 Spot 1 longitude ($^{\circ}$) & $164 \pm 5 $ & $171 \pm 6 $ & $164$ & $192 \pm 7 $  & adjusted \\
 Spot 1 latitude ($^{\circ}$) &  $39 \pm 2 $ & $35 \pm 2 $ & $39$ & $39 \pm 2 $  & adjusted \\
 Spot 1 radius ($^{\circ}$) & $24 \pm 1 $ & $29 \pm 1 $ & $24$ & $27 \pm 1 $  & adjusted \\
 Spot 1 temp. factor & $0.90$ & $0.90$ & $0.90$ & $0.85$  & adopted  \\
 Spot 2 longitude ($^{\circ}$) &  $-$ & $60 \pm 6 $ & $-$ & $-$  & adjusted \\
 Spot 2 latitude ($^{\circ}$) &  $-$ & $38 \pm 5 $ & $-$ & $-$  & adjusted \\
 Spot 2 radius ($^{\circ}$) &  $-$ & $13 \pm 1 $ & $-$ & $-$  & adjusted \\
 Spot 2 temp. factor & $-$ & $0.90 $ & $-$ & $-$  & adopted \\

\noalign{\smallskip}
 \hline
 \noalign{\smallskip}
 Relative radii $(r/a)$ &  \\
 \noalign{\smallskip}
 \hline
 \noalign{\smallskip}
 $r_1$ & $0.0062 \pm 0.0014$  & $0.0059 \pm 0.0014$ & $0.0062$  & $0.0066 \pm 0.0015$ \\
 $r_2 (volume-averaged)$ & $0.3088 \pm 0.0040 $ & $0.3108 \pm 0.0040 $ & $0.3088$ & $0.3095 \pm 0.0040 $ \\
 
 $r_2(pole)$  & $0.2975 \pm 0.0057$ & $0.3352 \pm 0.0058$ & $0.2975$ & $0.2982 \pm 0.0056$ \\
 $r_2(point)$ & $0.3323 \pm 0.0039$ & $0.3084 \pm 0.0039$ & $0.3323$ & $0.3334 \pm 0.0037$ \\
 $r_2(side)$  & $0.3064 \pm 0.0034$ & $0.2993 \pm 0.0035$ & $0.3064$ & $0.3072 \pm 0.0033$ \\
 $r_2(back)$  & $0.3209 \pm 0.0047$ & $0.3233 \pm 0.0047$ & $0.3209$ & $0.3218 \pm 0.0045$ \\

 \noalign{\medskip}
 \hline
 \end{tabular}

 \begin{list}{}{}
\item[$^{\mathrm{a}}$] Geometric parameters are fixed to those for the \textit{g} band, $T_1$ was adjusted to fit the total eclipse.
\item[$^{\mathrm{b}}$] $L_1,L_2$ - relative luminosities of the components

\end{list}

\end{table*}

\begin{table}
\centering
 \caption{Absolute parameters}
  \label{tab_abspars}
 %\begin{tabular}{p{3cm}p{2.5cm}}
 \begin{tabular*}{0.8\columnwidth}{l @{\extracolsep{\fill}} c}
 \hline
\noalign{\smallskip}
 Parameter &   Value \\

 \noalign{\smallskip}
 \hline
 \noalign{\medskip}
 $M_1$\,(M$_{\odot}$)       &  $0.723$   \\
 $M_2$\,(M$_{\odot}$)       &  $0.640$ (fixed)   \\
 $a$ (R$_{\odot}$)         &  $2.082$  \\
 $R_1$\,(R$_{\odot}$)       &  $0.013 \pm 0.003$  \\
 $R_2$\,(R$_{\odot}$)       &  $0.645 \pm 0.012$ \\
 $T_1$\,(K)               &  $8700 \pm 1100$  \\
 $T_2$\,(K)               &  $4070$ (fixed) \\
 $L^{bol}_1$\,(L$_{\odot}$) &  $(8.7 \pm 0.2) \times 10^{-4}$   \\
 $L^{bol}_2$\,(L$_{\odot}$) &  $0.103 \pm 0.003$ \\
 $log(g)_1$              &  $8.07 \pm 0.21$  \\  
 $log(g)_2$              &  $4.66 \pm 0.02$ \\  
 \noalign{\medskip}
 \hline
 \end{tabular*}
\end{table}

The analysis was performed with our own code for synthesis of light and radial velocity curves of close binary systems (CBSs) in the Roche model, which is similar to the well-known algorithm of Wilson and Devinney \citep{1971ApJ...166..605W, 1979ApJ...234.1054W} widely used in CBS studies. Our code is described in detail in \cite{1988SvA....32..608A, 1996ARep...40..483A} and \cite{2000ApJ...529..463A}. Here we describe its main features only. The computer code allows one to calculate light and radial velocity curves simultaneously, either for a circular or an eccentric orbit. Axial rotation of the components may be non-synchronized with the orbital revolution. Tidal and rotational distortion of the components, eclipses and possible spots on the surface of the components are taken into account. The intensity of the radiation coming from an elementary area of the stellar surface and its angular dependence are determined by the temperature of the star, gravitational darkening, limb darkening, and heating by radiation from the companion. The degree of filling the inner critical Roche lobe by a component is parametrizied by the so called Roche lobe filling factor $\mu$, If a component completely fills its critical Roche lobe, $\mu=1$. If the component underfills the critical Roche lobe, $\mu<1$.

Based on the available information on the system, we set some input parameters of the model. As stated above, from the available spectral data, the estimated spectral type of the MS star is K7V. Its temperature is $T_2\approx 4070$\,K (Fig.\ref{fig_eclipse}). The average mass of K7V stars is $M_2\approx 0.64$\,M$_{\odot}$ \citep{2012ApJ...757..112B}. The semi-amplitude of the MS star radial velocity curve from our spectral data is $K_2 = 178$\,km\,s$^{-1}$, which results in the mass function $f_2(m)=0.174$\,M$_{\odot}$. Our preliminary modelling has shown that the orbital inclination angle is $i\approx 73^{\circ}$. Thus, by using the third Keplerian law and the above value of the mass function, one can estimate the WD mass as $M_1\approx 0.723$\,M$_{\odot}$ and the mass ratio as $q=M_2/M_1\approx 0.885$. In further analysis, $q$ was set at this value. We also used the fixed values of the gravitational darkening coefficients $\beta_1=0.25$ \citep{1924MNRAS..84..702V} and $\beta_2=0.08$ \citep{1967ZA.....65...89L} as well as the albedos $A_1=1$ and $A_2=0.5$ \citep{1969AcA....19..245R}.

The synthetic light curves were computed in \textit{g}, V, \textit{r}, and \textit{i} bands and fitted to the observed data. The limb darkening of the K star was accounted for by non-linear laws: the logarithmic law for \textit{g}, V bands and the "square root" law for \textit{r}, and \textit{i} bands \citep{1993AJ....106.2096V}. The limb darkening law and its coefficients of the WD were taken from \cite{2013ApJ...766....3G}. The orbit was considered to be circular and the components rotation - synchronous with the orbital one. The orbital phases of the observed light curves were computed with our photometric ephemeris ${\rm HJD} = 2458762.2285 + 0.297549$~days. At the orbital phase $0.0$ the secondary component K7V is in front of the primary component (WD).

\begin{figure*}
  \includegraphics[width=0.8\textwidth]{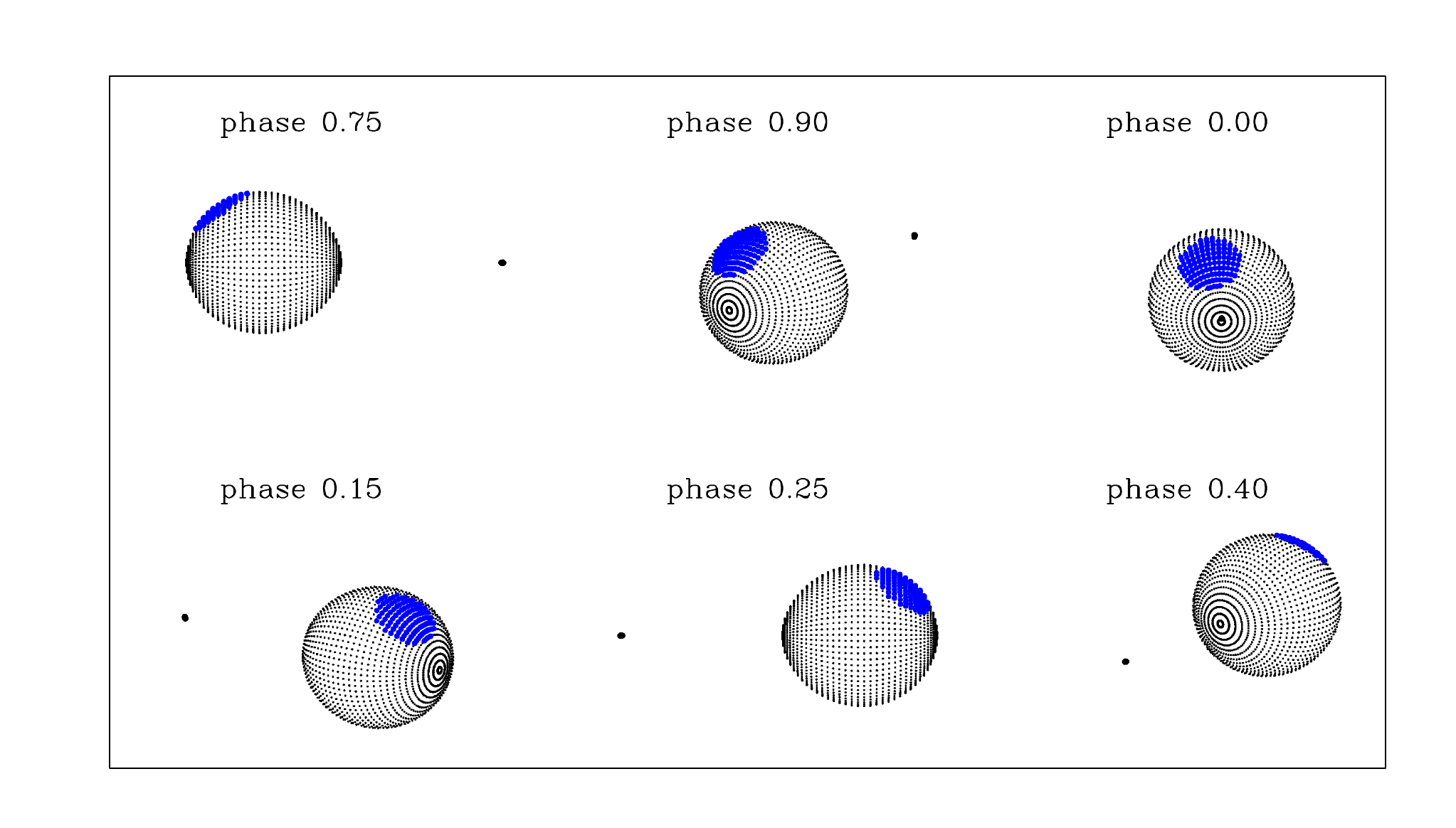}
  \caption{Model view of the binary system at the different orbital phases showing the visibility of the spot on the surface of the K star.}
    \label{fig_cartoon}
\end{figure*}

The variable parameters of the model are: the orbital inclination $i$, the Roche lobe filling factors $\mu_1, \mu_2$ and the temperature of the primary component $T_1$. To fit the asymmetry of the light curve we had to add one or two spots on the K star surface with the longitude, latitude, angular radius and temperature factor as their parameters. The temperature factor is the ratio of the spot temperature to the temperature of the underlying unspotted surface (for more details of spot modelling see \citealt{2017ApJ...835..251W}). As a rule, when fitting spot models, a fixed temperature factor is used \citep{2017ApJ...835..251W, 2019AJ....157....3L}. Similarly to these papers, we fixed the temperature factor at 0.85 and 0.9. The difference between the spot and stellar surface temperatures is then $\sim 600$\,K and $\sim 400$\,K respectively, which corresponds to the data from \cite{2005LRSP....2....8B} for K stars. The well-known Simplex algorithm \citep{10.1093/comjnl/7.4.308, 1987ApJ...313..346K} was used when fitting the model light curve to the data. The results of the fitting are shown in Tables~\ref{tab_modelpars},~\ref{tab_abspars} and Figs.~\ref{fig_lcfit} and~\ref{fig_cartoon}. The uncertainties provided in Tables~\ref{tab_modelpars} and~\ref{tab_abspars} are $1-\sigma$ intervals where the true parameter values are expected to be.

In the models in \textit{g}, V, \textit{i} bands two model parameters are reliably defined: the orbital inclination $i \approx 73\fdg1$ and the K star Roche lobe filling factor $\mu_2 \approx 0.86$ (Table~\ref{tab_modelpars}). Thus, the K star shape is tidally deformed, but the star does not fill its critical Roche lobe. This means that at the current evolutionary stage mass transfer from the K star to the WD through the inner Lagrangian point is possibly absent. 

Due to the asymmetry of the light curves (Fig.~\ref{fig_lcfit}) varying between the photometric bands the modelling was rather complicated. Note that in the \textit{g}, \textit{r} and \textit{i} bands observations were obtained during the same night, while the data in the V band was obtained a few days earlier. Adding one spot was sufficient to fit the light curve in the \textit{g} band. To fit the V curve, we had to add a small second spot on the surface of the K star. Note that {\bf g, V} model light curves are still not in perfect agreement with the observed one with either one or two spots. The small remaining differences are possibly related to the complex spot activity of the red dwarf (for an example, see \citealt{2016MNRAS.458.2793P}). We did not try to further improve the fit as it would require many more free parameters and make their values rather ambiguous. Also, the spot activity is variable over large time scale (see below). The main objective of our work was to estimate the main orbital and component parameters. Due to a unique combination of the ellipsoidal and eclipsing variability, the geometric parameters and the WD temperature can be indeed found with high accuracy, as even a slight change in these parameters leads to strong discrepancies between the model and the observed light curves.

The \textit{r} light curve is significantly different from all the others, the main differences are that the maximum at the phase $\sim 0.25$ is higher than the one at the phase $\sim 0.75$, and the eclipse is significantly deeper than that in the other bands. The out-of eclipse light curve cannot be fitted by a combination of spots on the surface of the K star, as it is highly unlikely that such spots would only demonstrate themselves in the \textit{r} band alone. Moreover, no spots can explain the deep eclipse in this band. It is likely that the unusual shape of the \textit{r} light curve is related to the properties of the WD. We discuss these matters below in Section \ref{sec_disc}. Thus, when computing the \textit{r} model light curve, we fixed geometric parameters to those found in the \textit{g} band and fitted $T_1$ only.

In the \textit{i} band, the solution is similar to the \textit{g} band, although small systematic differences between the model and the data are present in the phase interval 0.05-0.45. Adding a second spot does not improve the fit. Also, the eclipse depth is larger than it should be according to $T_1$ found in the bands \textit{g} and V. We conclude that the differences in the \textit{i} band compared to \textit{g} and V are probably of the same nature as in the \textit{r} band, although of smaller magnitude.

Overall, based on our modelling one can conclude that out of eclipse variability of the system is mainly determined by tidal deformation of the K star and by the presence of a low temperature spot(s) on its back half (the back half of the star is opposite to the Lagrangian point $L_1$ and is visible to the observer in the orbital phase interval (approximately $0.75-0.40$, see Fig.~\ref{fig_cartoon}). The reflection effect at the surface of the K star is minor, as the WD temperature is limited by $T_1 \le 10\,000$~K (from our solution in the \textit{g}, V bands as well as from {\em GALEX} FUV data).   

The narrow total eclipse of the WD allowed us to estimate its temperature. However, as said above, it was impossible to successfully fit the eclipse in all four bands with the same temperature. In the \textit{g} and V bands the best fit is obtained with the WD temperature $T_1 = 8700\pm 1100$\,K (Fig.~\ref{fig_lcfit}, right panel). However, in the red bands the WD temperature must be increased. While in the \textit{i} band the increase is moderate ($T_1 \sim 14\,000 $~K), in the \textit{r} band it is much larger ($T_1 \sim 22\,000 $ K). Such high values of temperature (especially $T_1 \sim 22\,000 $~K in the \textit{r} band) contradict the UV data which suggest that the WD temperature should be smaller than $10\,000$~K.

\subsection{Light curves slow variability}\label{sec_LC_var}
We also detected slow out-of-eclipse variability of the light curve in the data gathered by the MASTER telescope simultaneously in V and R bands from January 2019 to April 2019 during 160 periods of GPX-TF16E-48. Observed variations in V and R were correlated in time. Recent additional observations in \textit{SDSS g}, \textit{r} and \textit{i} filters on 2019 October 7 confirmed the variability of the light curve of the object. The changes in the elliptical variability of GPX-TF16E-48 can be explained by variations of the spot covering fractions and/or temperatures of spots on the surface of the MS companion. However, notable decrease of the depth of the eclipse in \textit{SDSS r} and \textit{i} filters can be related only to the properties of the eclipsed body (right panel of Fig.~\ref{fig_eclipse}).

\section{Discussion}\label{sec_disc}

Independently of the complexities described in the previous section, the orbital inclination $i$ and Roche lobe filling factors $\mu_1$, $\mu_2$ were defined with high accuracy. The ellipsoidal variability is mainly determined by a combination of $i$ and $\mu_2$. At the same time, the narrow total eclipse of the WD depends on $i$, $\mu_1$, $\mu_2$, $T_1$. 

The abnormal shape of the \textit{r} (and to a smaller degree, \textit{i}) bands requires special discussion. As stated above, the differences in the out of eclipse light curve shape in \textit{r} and other bands cannot be explained by any spots on the surface of the K star. Such spots would demonstrate themselves in all bands. Moreover, any such spots cannot explain the increased depth of the eclipse. Note that in the \textit{i} band, the eclipse is also deeper than it should be at the WD temperature found from the \textit{g} and V data, and the out of eclipse light curve is also deviating from the model. It seems likely that these features are related to the properties of the WD. In our model the WD temperature $T_1$ is just a measure of the WD luminosity (at a given WD radius). One could suggest that the actual temperature is that found from \textit{g} and V data, and that some additional emission of the WD is present, with the majority of the corresponding flux occurring in the \textit{r} band but also demonstrating itself in the \textit{i} band. It can be assumed that the excess in \textit{r} is caused by H$\alpha$ emission, but we did not detect strong enough lines in the spectra. Moreover, the depth of the eclipse in the H$\alpha$ filter does not exceed the depth in R and \textit{SDSS r} (left panel of Fig.~\ref{fig_eclipse}). These facts suggest that the source of excess should have rather wide spectral features.
One possible explanation could be that the WD has a strong magnetic field coupled with very low-accretion rate of stellar wind from the secondary star (the so called LARP object). These could result in cyclotron emission centered around the \textit{r} band and having rather wide spectral distribution (\citealt{2001A&A...374..189S},  \citealt{2002ASPC..261..102S}, \citealt{2005ApJ...630.1037S}, \citealt{2005ASPC..330..137W}). Such a mechanism would allow us to explain the unusual spectral energy distribution of the WD (right panel of Fig.~\ref{fig_eclipse}) (following from the depths of the eclipses in various bands) and the out of eclipse variations in the \textit{SDSS i} and \textit{r} bands. This hypothesis has an additional circumstantial evidence: significant variations of brightness of the WD on a short timescale and clear signs of strong chromospheric activity of the K star.
To check this hypothesis, high S/N spectra with good time resolution and polarimetric observations are necessary. Until then, trying to fit the observed light curves in the \textit{r} and \textit{i} bands would be too speculative.

Obtained system parameters allowed us to make some assumptions about the past and future of the GPX-TF16E-48. Based on the current mass of the WD and the semi-empirical MIST model of initial-final mass relation of WDs from \citealt{2018ApJ...866...21C}, the progenitor mass was close to $2.9$\,M$_{\odot}$. After the main sequence stage, the primary evolved into a red giant forming a common envelope and then into a cooling white dwarf. Using WD cooling tracks for different masses from \citealt{1997ApJ...486..413S} and \citealt{1995LNP...443...41W} (shown in Fig. 7 of \citealt{2003A&A...406..305S}), we estimated an age of the WD as 1 Gyr. According to \citealt{1981A&A...100L...7V}, \citealt{1983ApJ...275..713R} and \citealt{2003A&A...406..305S}, for the system with $M_2>0.3$\,M$_{\odot}$ a dominant mechanism of orbital angular moment loss is a magnetic breaking. With this assumption, we can assess the orbital period at the end of the common envelope phase using Eq. (9) from \citealt{2003A&A...406..305S} as $P_{ce}\approx 1.05$\,d. Due to further loss of orbital angular momentum of the system, the separation between the components will decrease and Roche lobe of secondary MS star will shrink. When the effective radius of Roche lobe reaches the radius of secondary, the system will become semi-detached, and  stable mass transfer from the MS to the WD will begin. According to \citealt{1983ApJ...268..368E}, and for the obtained system parameters, we derive an orbital separation and period at the beginning of semi-detached phase as $a_{sd}\approx  1.75$\,R$_{\odot}$ and of $P_{sd}\approx 5.5$\,hours respectively. The time before the onset of the semi-detached phase calculated by Eq. (9) from \citealt{2003A&A...406..305S} is $t_{sd}\approx 8.9 \times 10^6$\,years.

\section{CONCLUSIONS}\label{CONCL}
The discovery and the results of the follow-up observations and preliminary light curve modelling of the bright eclipsing post-common envelope binary GPX-TF16E-48 are presented. It is shown that the system consists of a cool white dwarf (WD) and a K7V star. 

Based on our analysis of the spectral and photometric data, we estimated the parameters of the system. The K7V star is tidally deformed but does not fill its Roche lobe ($\mu_2\approx 0.86$). The orbital inclination $i=73^\circ.1\pm 0^\circ.2$ and the mass ratio $q=M_2/M_1\approx 0.88$. The parameters of the K7V star are $M_2\approx 0.64$\,M$_{\odot}$ (adopted from the average mass of K7V stars, \citealt{2012ApJ...757..112B}), $R_2=0.645\pm 0.012$\,R$_{\odot}$, and $T_2\approx 4070$\,K. Note that the obtained radius of the K7V star is close to the typical radius ($0.65$\,R$_{\odot}$) from \cite{2012ApJ...757..112B}. The parameters of the white dwarf are $M_1\approx 0.72$\,M$_{\odot}$, $R_1=0.013\pm 0.003$\,R$_{\odot}$,  $T_1=8700\pm 1100$\,K.

The depth of eclipse in various bands revealed unusual spectral energy distribution of the eclipsed body. In the case of an eclipsing WDMS system, the depths of the eclipse observed in different bands vary and they are determined by the ratio of temperatures of the components. For a typical system, which consists of a late red dwarf and a WD, observed eclipse depths increase from IR to UV \citep{2004MNRAS.355.1143M, 2012ApJ...757..133L, 2013MNRAS.436..241P, 2019AJ....157....3L}. However, in the case of GPX-TF16E-48, multi-colour photometric follow-up observations revealed that depths of eclipses show an unusual colour dependence with an excess in red filters (see Fig.~\ref{fig_RV_LC} and Fig.~\ref{fig_eclipse}). Also, a change of the depth of total eclipse was detected, which occur in a monthly timescale.

We suspect that the primary (WD) companion of GPX-TF16E-48 has a strong magnetic field and accrete stellar wind from the secondary companion at a low rate. These could result in cyclotron emission, which dominates in the observed range of the WD spectrum.
Additional photometric, spectroscopic and polarimetric observations with high time resolution are necessary to determine the real nature of the GPX-TF16E-48. The object will be observed by TESS \citep{2014SPIE.9143E..20R} in Sector 18 from 2019 November 02 to 2019 November 27 (TIC 353075202, TESS mag\,=\,13.15). Though TESS data is gathered in just one filter, it will be useful to look for possible eclipse timing and shape variations, the spot and chromospherical activity of the K7V companion.

\section*{Acknowledgements}
\label{Ack}
This work has made use of data from the European Space Agency (ESA)
mission {\it Gaia} (\url{https://www.cosmos.esa.int/gaia}), processed by
the {\it Gaia} Data Processing and Analysis Consortium (DPAC,
\url{https://www.cosmos.esa.int/web/gaia/dpac/consortium}). Funding
for the DPAC has been provided by national institutions, in particular
the institutions participating in the {\it Gaia} Multilateral Agreement.

This research has made use of the NASA/IPAC Extragalactic Database (NED), which is operated by the Jet Propulsion Laboratory, California Institute of Technology, under contract with the National Aeronautics and Space Administration.

The work of IA (light curves analysis) was supported by the RSF grant 17-12-01241. The work of EA was supported by the Program of development of M.V. Lomonosov Moscow State University (the Leading Scientific School ''Physics of stars, relativistic objects and galaxies''). OB thanks (T\"{U}B\.{I}TAK) for partial support in using T100 telescope with project numbers 19AT100-1346. AB would like to thank Dr.~Amaury Triaud for helpful discussions about the system. TRAPPIST-North is a project funded by the University of Liege, in collaboration with Cadi Ayyad University of Marrakech
(Morocco). TRAPPIST is a project funded by the Belgian Fonds (National) de la Recherche Scientifique (F.R.S.-FNRS) under grant FRFC 2.5.594.09.F. E.J and M.G are F.R.S.-FNRS Senior Research Associates. The INT is operated on the island of La Palma by the Isaac Newton Group of Telescopes in the Spanish Observatorio del Roque de los Muchachos of the Instituto de Astrof\'isica de Canarias. The work of VK and AP was partially supported by the Russian Ministry of Science and Education, FEUZ-2020-0030.
%%%%%%%%%%%%%%%%%%%%%%%%%%%%%%%%%%%%%%%%%%%%%%%%%%



\bibliographystyle{mnras}
\bibliography{bibliography}
%must be included in an alphabetical list
%\begin{thebibliography}{99}

%\end{thebibliography}
%%%%%%%%%%%%%%%%%%%%%%%%%%%%%%%%%%%%%%%%%%%%%%%%%%

%%%%%%%%%%%%%%%%% APPENDICES %%%%%%%%%%%%%%%%%%%%%

%%%%%%%%%%%%%%%%%%%%%%%%%%%%%%%%%%%%%%%%%%%%%%%%%%

% Don't change these lines
\bsp	% typesetting comment
\label{lastpage}
\end{document}